\documentclass[preprint,12pt]{elsarticle}

\usepackage{amssymb}
\usepackage{amsthm}
\usepackage{mathtools}
\usepackage{subfigure}
\usepackage{float}
\providecommand{\e}[1]{\ensuremath{\times 10^{#1}}}

\journal{-}

\begin{document}

\begin{frontmatter}



\title{Phase-Space Analysis of Fluid-Dynamical Models of Traffic Flow}

\author[rvt]{M David Alvarez Hernandez\corref{cor1}}
\ead{mdalvarezh@xanum.uam.mx}
\author[focal]{Shani Alvarez Hernandez}
\ead{seah@xanum.uam.mx}

\cortext[cor1]{Corresponding author}

\address[rvt]{Department of Physics, Universidad Autonoma Metropolitana, San Rafael Atlixco No. 186, Col. Vicentina, Iztapalapa, 09340, Mexico.}

\address[focal]{Department of Mathematics, Universidad Autonoma Metropolitana, San Rafael Atlixco No. 186, Col. Vicentina, Iztapalapa, 09340, Mexico.}



\begin{abstract}
The present research article is devoted to the study of two fluid-dynamical traffic models and their steady states by first analyzing the kinetic-type Borsche-Kimathi-Klar model and the Navier-Stokes-type Helbing model. To fulfill this purpose, the study is made by means of a general analysis in the phase space. Phase space paths are constructed in several cases for both models, evaluating explicitly some phase space trajectories. Ultimately, the results of these analyses are discussed.

\end{abstract}


\end{frontmatter}


\section{Introduction}
\label{sec:1}

The year 2008 made manifest a defining moment in the complex and ongoing urban revolution; for the first time in history, more than 50 percent of the world's population was based in urban areas, with the current urbanization rate in such a state that, if it keeps holding, the urban share of the global population could reach 60 percent by 2030, according to UN projections. It is the case that in more developed regions, the number of people living in urban areas will rise only slightly in the next 20 years, while the less developed regions will experience a particularly sharp and increasing rate. As such, one of the main concerns in modern urban planning is the providing of good and efficient transport infrastructure. \cite{salyer}\\

Consequently, it is no surprise that in recent decades scientists have tried to develop mathematical models of traffic flow with the purpose of studying, planning and optimizing transport infrastructure and their vehicle flow. Traffic flow models constitute a theoretical framework in which to study the complex behavior found in different traffic scenarios, and its origins can be traced back to two main lines of study: the microscopic and macroscopic approach.  Apart from some follow-the-leader models and some microscopic models of driver behavior, most of the proposed models make use of the analogy between vehicular traffic and gases or fluids; gas kinetic models, cellular automaton models, and fluid-dynamical models have been suggested. Some authors have reviewed the state-of-the-art concerning these methodologies. \cite{helbing01}\\

Fluid-dynamical models are derived from fluid-dynamical equations, which are in close analogy to a Navier-Stokes viscous fluid model. Among the most handled fluid-dynamic models is the second order Kerner-Konh\"{a}user model, which has been considered a classic instance since it contains focal features of the traffic problem, such as the fundamental diagram $V_e$, which represents the correlation between the average velocity and the vehicle density (or flux and density), and the values of the relaxation time $\tau$, representing the time span in which drivers adjust their velocity to that of $V_e$. Both are taken accordingly to certain experimental data.\\ 

With the aim to solve some of the problems presented by the Kerner-Konh\"{a}user model \cite{helbing95}, several augmented models have been proposed. In this work we will focus our study on two amplified fluid-dynamic models: a modified Borsche-Kimathi-Klar (BKK) model and a modified Helbing model. In both models the vehicle density and the average velocity are taken as the main relevant variables for vehicles in a one-lane highway with no ramps.\\

The Helbing model can be considered an enhanced version of the classical Kerner-Konh\"{a}user model. This model proposes as dynamical variables the vehicle flow velocity $V(x,t)$, the vehicle density $\rho(x,t)$ and the velocity variance $\Theta(x,t)$. The model uses the two main equations of the Kerner-Konh\"{a}user model and introduces a third equation for the velocity variance of vehicles. In addition, the model considers as parameters the average vehicle size and the response drive time.\\

On the other hand, the BKK model was developed by taking into consideration the kinetic models proposed in [11, 16, 18, 19] but from a fluid-macroscopic point of view, observing that the derived macroscopic equations contain Hamilton-Jacobi terms. The model advocates as dynamic variables the average vehicle density $\rho(x,t)$ and the average vehicle flow velocity $V(x,t)$. The description of both variables is given by a continuity equation for the vehicle density, and a dynamic equation for the velocity.\\

Both models admit solutions for a homogeneous steady state, however, some of the latest research, see [8], have suggested an appropriate methodology to go from the partial differential equations of the model to ordinary differential equations by means of a change in the reference frame. Therefore, the traffic model equations can be studied with a sensible physical meaning while the theory of dynamical systems can be used to understand the results. Phase space analysis provides us with an excellent tool to achieve this goal. The general characteristics of traffic models, such as their nonlinearity, indicate that this kind of analysis can offer a better understanding of the dynamics of traffic flow.\\

We will consider the dynamic portion as seen from a moving reference frame which travels with constant velocity $V_g$, thus changing the reference frame and allowing us to obtain a new set of ordinary differential equations for both models. Next, we will look for steady states in this new reference system, see \cite{kerner94} and \cite{velasco11}, which can be studied in the phase space, thus identifying the orbits in the system and possible heteroclinic trajectories.\\

In light of this, the purpose of this work is to apply the previously decribed systematic study of the steady states of the modified BKK and Helbing models in order to study each separately, and ultimately compare them. For this purpose, a viscosity term and the fundamental diagram is added in the BKK model to simulate the structure of a Navier-Stokes equation. As for the Helbing model, the variance is taken as a function of the velocity to close the system, thus shaping the variance from the start so that the only dynamic variable will be the velocity. This last change is necessary if a qualitative comparison between the models is to be made. As it is presented, a comparison is not straightforward because of the different order of the equations found in both models.\\

The analysis is organized as follows. In section 2, a general scheme of the study is introduced; section 3 will be dedicated to the modified BKK model analysis; section 4 will examine the modified Helbing model; in section 5 we present some examples of numerical simulations in both models, before concluding the examination with a comparison between the models plus some final remarks.


\section{General Scheme}
\label{sec:2}

From a general perspective, let us first consider the equations of motion for representative macroscopic, fluid-dynamical traffic models. This means that the following section will be devoted to models in which we can find a set of partial differential equations, such as the Navier-Stokes type of equations for macroscopical quantities, e.g. vehicle density $\rho(x, t)$ or the average vehicle velocity $V(x,t)$. As mentioned before, the common characteristic found in these models is that they share the continuity equation for density and the structure of the equation of motion for macroscopic quantities.\\

Let us assume specifically for both models a one-dimensional highway with no ramps, i.e. no flow sources, in such a way that the total number of vehicles in the highway remains constant. Then, the equations for the chosen models can be written in the so-called conservative form

\begin{equation}\label{Eq1RMV}
	\frac{\partial \phi}{\partial t} + \frac{\partial f}{\partial x} = S,
\end{equation}

\noindent
where the functions $\phi$ and $f$ are given by

\begin{equation}\label{Eq2RMV}
\phi = 
 \begin{pmatrix}
  \rho \\
  \rho V
 \end{pmatrix} , \qquad f = 
 \begin{pmatrix}
  \rho V\\
  \rho V^{2} + \mathcal{P} 
 \end{pmatrix}
\end{equation}

It can be noted that the source term in Eq.(\ref{Eq1RMV}) vanishes as a consequence of the choice of system conditions, due to the absence of ramps and exits. This means that the average vehicle density $\rho(x,t)$ satisfies the continuity equation

\begin{equation}\label{Eq4RMV}
	\frac{\partial \rho}{\partial t} + \frac{\partial}{\partial x} (\rho V) = 0.
\end{equation}

The dynamics of the system can be explored from a reference frame moving at a constant velocity $-V_g$, in such a way that the next change of variable can be used
	
\begin{equation}\label{Eq5RMV}
\xi = x + V_{g}t
\end{equation}

\noindent 
and Eq.(\ref{Eq4RMV}) can be integrated directly, resulting in

\begin{equation}\label{Eq6RMV}
\rho(V_{g} + V) = Q_{g}, \qquad \rho = \frac{Q_{g}}{(V_{g} + V)},
\end{equation}

\noindent 
wherein $Q_g$ is a constant of integration. The direct application of the change of variables in the equations of motion specified by Eq.(\ref{Eq1RMV}) will lead to new second order differential equations, which will have the following general structure

\begin{equation}\label{Eq8RMV}
M\dfrac{d^{2}V}{d\xi^{2}} + D_{1}\dfrac{dV}{d\xi} + D_{2}\biggl(\dfrac{dV}{d\xi}\biggr)^{2} = F
\end{equation}

The coefficients $M, D_{1}, D_{2}$, and $F$ are functions of $(V; \Theta, V_{g}, Q_{g})$, and each one will be specified for every case model. In all cases $V=V(\xi)$ is the unknown dynamic variable.\\ 

From the point of view of one-particle dynamics, the former equation could be interpreted as the equation of motion for a particle of mass $M$ in a Stokes-type fluid, where a force field $F$ is present and the friction terms, measured by the coefficients $D_{1}, D_{2}$, are also present. The force field could be obtained from a fictional potential $U(V)$ in such a way that 

\begin{equation}\label{potential}
F = - \dfrac{dU}{dV},
\end{equation}

\noindent
where the potential $U$ can be also a function of $(V; \Theta, V_{g}, Q_{g})$. The critical points of this potential, which must comply the condition $F=0$, and the sign of the derivative of {\footnotesize $\dfrac{dF}{dV} = -K$} at the critical points will tell us if the potential has reached a maximum or a minimum. The stability characteristic of these critical points will determine the global dynamics of the traffic flow.\\

To keep the analyses as simple as possible, the next set of dimensionless variables are defined, see \cite{velasco11} and \cite{velasco12},

\begin{equation}\label{Eq9RMV}
z=\rho_{max}\xi, \quad v= \dfrac{V}{V_{max}}, \quad v_{g}=\dfrac{V_{g}}{V_{max}}, \quad q_{g}=\dfrac{Q_{g}}{\rho_{max}V_{max}}, \quad \theta=\dfrac{\Theta}{(V_{max})^{2}}.
\end{equation}

\noindent
So Eq.(\ref{Eq8RMV}) can be rewritten in dimensionless terms as follows

\begin{equation}\label{Eq10RMV}
\dfrac{d^{2}v}{dz^{2}} + d_{1}(v;\theta, v_{g}, q_{g})\dfrac{dv}{dz} + d_{2}(v;\theta, v_{g}, q_{g})\biggl(\dfrac{dv}{dz}\biggr)^{2} = f(v;\theta, v_{g}, q_{g}),
\end{equation}

\noindent
in which the dimensionless pseudo-friction coefficients $d_1$, $d_2$, and the pseudo-force $f$, are defined as

\begin{equation*}\label{EqsnRMV}
d_{1}(v;\theta, v_{g}, q_{g})=\dfrac{D_{1}}{M}, \quad d_{2}(v;\theta, v_{g}, q_{g})=\dfrac{D_{2}}{M}, \quad f(v;\theta, v_{g}, q_{g}) = \dfrac{F}{M}.
\end{equation*}

A qualitative description of the system described by Eq.(\ref{Eq10RMV}) can be made at this point. In order to perform the system's phase-space analysis, the order of the differential equation is lowered to first order by applying the transformations
	
\begin{equation}\label{Eq11.0RMV}
\frac{dv}{dz} = c_{1} = y,
\end{equation}	

\begin{equation}\label{Eq11.1RMV}
\dfrac{dy}{dz} = c_{2} = - d_{1}y - d_{2}y^{2} + f
\end{equation}.

\noindent
The properties of the dynamical system defined in Eqs.(\ref{Eq11.0RMV}, \ref{Eq11.1RMV}) are determined by the functions $c_1$, $c_2$, and it is clear that the critical points can be calculated from

\begin{equation}\label{Eq12RMV}
c_{1}(v_{c}; \theta, v_{g}, q_{g})=0, \qquad c_{2}(v_{c}; \theta, v_{g}, q_{g})=0
\end{equation}
\\
\noindent
wherein $v_c$ corresponds to the dimensionless velocity at the critical points, with coordinates in phase space $CP = (v_{c}, 0)$; it should be noted that their values must be in the range $[0,1]$ for the chosen values of $-1 \leq v_{g} \leq1$ and $-1 \leq q_{g} \leq 1$. There can be also several critical points, since the quantity $c_{2}(v_{c}; \theta, v_{g}, q_{g})$ could have several roots.\\

For the purposes of the system's stability analysis, it will be assumed that the function $f$ has the following structure

\begin{equation}
f = [v_{e}(v_{c})- v_{c}]
\end{equation}
\\
\noindent
where we will consider two different cases for $v_{e}(v_{c})$, which represents the velocity's stationary state. In the first case, $v_{e}(v_{c})$ is expressed as the Greenshields fundamental diagram

\begin{equation}\label{Greenshields}
v_{e}(v_{c}) = \dfrac{(v_{c} + v_{g}) - q_{g}}{v_{c} + v_{g}}
\end{equation}
\\
\noindent
and as for the second case, $v_{e}(v_{c})$ is expressed as the Kerner-Konh\"{a}user diagram

\begin{equation}\label{KernerK}
v_{e}(v_{c})=\left[\left(1+e^{\left(\frac{\frac{q_{g}}{v_{c} + v_{g}} - 0.25}{0.06}\right)}\right)^{-1} \right] - 3.72\e{-6}
\end{equation}
\\

The system's stability characteristics can be determined by the linearized dynamical system around each critical point. For that, the system's Jacobian matrix and their associated eigenvalues need to be calculated

\begin{equation}\label{jacobian}
J(CP) = \begin{pmatrix}
\partial_{v}(c_{1})_{CP} & \partial_{y}(c_{1})_{CP} \\
\partial_{v}(c_{2})_{CP} & \partial_{y}(c_{2})_{CP} 
\end{pmatrix} = \begin{pmatrix}
j_{1,1}(CP) & j_{1,2}(CP)\\
j_{2,1}(CP) & j_{2,2}(CP)
\end{pmatrix}.
\end{equation}
\\
\noindent
A straightforward calculation shows that 

$$j_{1,1} = 0, \quad j_{1,2} = 1, \quad j_{2,1} = \biggl(\dfrac{\partial f}{\partial v} \biggr) = -K(v_{c}; \theta, v_{g}, q_{g}), \quad j_{2,2} = - d_{1}(v_{c}; \theta, v_{g}, q_{g})$$
\\ 
\noindent
Where the eigenvalues of the Jacobian evaluated at each critical point are given by

\begin{equation}\label{eigenvalues}
E_{\pm}= \dfrac{-d_{1}(v_{c})}{2} \pm \sqrt{\biggl[ \dfrac{d_{1}(v_{c})}{2} \biggr]^{2} - K(v_{c}; \theta, v_{g}, q_{g})}.
\end{equation}

In order to find the local stability properties of the nonlinear system, we apply the Hartman-Grobman theorem that states that critical points of a nonlinear system whose corresponding Jacobian matrix has eigenvalues with real part different than zero, have the same behavior as the critical points of the linearized system. These critical points are called hyperbolic points, see \cite{jordansmith} and \cite{tu}.\\

It is worth mentioning that the linear stability analysis only gives a qualitative picture near the critical points. Moreover, the nonlinear term with respect to $y$ (the coefficient $d_{2}$) does not play a role in the linearized analysis, but in the simulation of orbits in the phase space for the complete nonlinear system, the term $d_{2}$ is relevant since the order of magnitude is the same as the linear contributions.\\

Returning to the analogy with the one-particle motion interpretation, we can notice that stable critical points correspond to minimum points of the pseudo-potential $U$. Also, a physical interpretation of the friction coefficient $d_{1}$ can be offered in terms of the system's total energy; $d_{1}$ is considered as the friction linear term, linear to $y$, while $d_{2}$ shall be considered as the nonlinear friction term.\\

\noindent
As noted in the one-particle analogy, the kinetic energy per unit of mass of the particle is given by \begin{footnotesize} $\dfrac{y^{2}}{2}$\end{footnotesize}, then from Eq.(\ref{Eq10RMV}) and using the fact that $f$ is derivable from the potential $U$, we get

\begin{equation}\label{energy}
\dfrac{d}{dz} \biggl(\dfrac{y^{2}}{2} + U \biggr) = -d_{1}y^{2} - d_{2}y^{3}
\end{equation}

\noindent
In Eq.(\ref{energy}) the terms $d_{1}y^{2}$ and $d_{2}y^{3}$ are a measure of the system's energy change rate, and their sign will determine if there exists generation or dissipation of the total energy in the dynamical system. The term $-d_{1}y^{2}$, which is positive when $d_{1} < 0$, indicates generation of energy, whereas when $d_{1} > 0$ there is dissipation. On the other hand, the contribution of the term $d_{2}y^{3}$ depends on the sign of the factors $d_2$ and $y^{3}$.\\

Therefore, the stability results found through the analysis of the jacobian's eigenvalues for each critical point should be consistent with the energy considerations expressed in Eq.(\ref{energy}). When the value of the parameters $q_g$ or $v_g$ makes an unstable critical point become stable, the system's total energy must be dissipated; otherwise, when a stable critical point becomes unstable then the system receives enough energy to escape.\\

The analysis concludes with the complete solution of the ordinary differential equations constituting the dynamical system, given by Eqs. (\ref{Eq11.0RMV}, \ref{Eq11.1RMV}). The solution of this set of equations is numerical and is represented qualitatively in the phase space. Since it cannot be done generally, each model must be studied separately.


\section{The modified Borsche-Kimathi-Klar Model}
\label{sec:3}

This model considers an underlying system of kinetic equations based on Fokker-Planck type of equations, from which a set of macroscopic equations are derived, see \cite{borsche}. This set of equations considers as dynamic variables the average spatial vehicle density $\rho(x,t)$ and the average traffic flow velocity $V(x,t)$. Both are defined respectively by

\begin{equation*}\label{BKK1}
\rho(x,t)=\int_{0}^{v_{max}}f(x,u,t)du
\end{equation*}

\begin{equation*}\label{BKK2}
V(x,t)=\int_{0}^{v_{max}}uF(x,u)du,
\end{equation*}
\\
\noindent
where $v_{max}$ denotes de maximal velocity allowed and $F(x,u)$ denotes the probability distribution in $u$ of cars at $x$, so $f(x,u,t)=\rho(x,t)F(x,u)$ denotes the average phase space density of vehicles with speed $u$ at place $x$ and time $t$. Then, the kinetic Fokker-Planck equation is given by, see \cite{illner}

\begin{equation}\label{BKK2.6}
\partial_{t}f + u\partial_{x}f = -\partial_{u}(B[f]f)=C^{+}(f).
\end{equation}

Here, $f$ stands once again for a traffic distribution function. Likewise, we denote by $\rho$, $u$ the macroscopic density and speed associated with $f$. From Eq. (\ref{BKK2.6}), by multiplying the kinetic equation by $u$ and integrating with respect to $u$, the next group of macroscopic equations is obtained

\begin{equation}\label{BKK3.1}
\frac{\partial \rho}{\partial t} + \frac{\partial (\rho V)}{\partial x} = 0
\end{equation}

\begin{equation}\label{BKK3.11}
\frac{\partial (\rho V)}{\partial t} + \frac{\partial}{\partial x}(\rho V^{2} + \mathcal{P}) - S = \frac{\rho}{\tau} [V_{e}(V) - V].
\end{equation}
\\
As stated before, the term $[V_{e}(V) - V]$, i.e. the fundamental diagram, is an \textit{ad hoc} addition not contemplated in the original model, but that must be considered in order to account for the empirical relation between flow velocity and density \cite{DSV}. This way, the model is comprised of two equations to describe the system: the continuity equation, seen in Eq.(\ref{Eq4RMV}), and the dynamical equation for the velocity. For this model, the traffic pressure $\mathcal{P}$ and source $S$ terms are given by

\begin{equation*}\label{BKK3.2}
\begin{split}
\mathcal{P} &= \int_{0}^{v_{max}} (u - V)^{2}f(x,u,t) du - \eta\dfrac{\partial V}{\partial x}\\
S &= \int_{0}^{v_{max}} uC^{+}(f)(x,u,t) du
\end{split}
\end{equation*}
\\
\noindent Following the argument presented in \cite{borsche}, the integral term on traffic pressure can be neglected, i.e. {\footnotesize ${P\approx - \eta\dfrac{\partial V}{\partial x}}$}, and the source term can be approximated as

\begin{equation*}
S \approx \rho b(\rho, V)\left|\frac{\partial V}{\partial x}\right| \frac{\partial V}{\partial x},
\end{equation*}

\noindent
where the coefficient $b(\rho, V)$ is given by
\vspace{-0.5cm}
\begin{center}
	\begin{equation}
	b(\rho, V) = \left\{
	\begin{tabular}{llll}
 $\dfrac{H_{B}^{2}P_{B}}{1/\rho - H_{B}}$	& & &  $\partial_{x}V < 0$\\
 $\dfrac{H_{A}^{2}}{1/\rho - H_{B}} \exp(-\tilde{\rho}(H_{A}-H_{B}))$	& & & $\partial_{x}V > 0$ \\ 
    \end{tabular}
    \right.
    \notag
   \end{equation}
\end{center}
\vspace{0.4cm}

This coefficient depends on the choice of the phenomenological braking density function $P_{B}$. An acceleration threshold $H_{A}$ and a braking threshold $H_B$ are also embedded in this coefficient, which are defined as

\begin{equation*}
H_{X}=H_{X}(V) = H_{0} + V\mathcal{T}_{X}, \qquad X= B, A
\end{equation*}
\\
\noindent
where  $H_A$ and $H_B$ are constants, $\mathcal{T}_{B} < \mathcal{T}_{A}$ are reaction times and $H_{0}$ denotes the minimal distance between the vehicles. A simplification of the coefficient can be enforced if we make $H_{A}=H_{B}=H$ and the breaking probability equal to one, i.e. $P_{B}=1$, thus obtaining

\begin{equation*}\label{BKK3.7}
b(\rho)= \dfrac{H^{2}}{1/\rho - H} = \dfrac{H}{1/\rho H - 1}
\end{equation*}
\\
\noindent
Therefore, the velocity equation for the modified BKK model is

\begin{equation}\label{Eq2BKK}
\frac{\partial (\rho V)}{\partial t} + \frac{\partial (\rho V^{2})}{\partial x} - \eta\dfrac{\partial^{2} V}{\partial x^{2}} - \biggl(\dfrac{\rho H}{1/\rho H - 1}\biggr) \left|\frac{\partial V}{\partial x}\right| \frac{\partial V}{\partial x} = \frac{\rho}{\tau} [V_{e}(V) - V]
\end{equation}
\\
\noindent
Now we proceed to employ the approach explained in Sect. \ref{sec:2}. Applying the change of variable proposed in Eq. (\ref{Eq5RMV}) to the continuity equation, Eq. (\ref{BKK3.1}), the equation now transforms into

\begin{equation}\label{Eq4pollo}
\rho(V) = \frac{Q_{g}}{(V + V_{g})}
\end{equation}
\\
\noindent
With the change of variable and the substitution of the density according to Eq.(\ref{Eq4pollo}), while taking into account the previously defined dimensionless variables, Eq.(\ref{Eq9RMV}), the dimensionless equation for the velocity is

\begin{equation}\label{Eq8pollo}
- \biggl(\dfrac{n(v + v_{g})}{q_{g}}\biggr)\dfrac{d^{2}v}{dz^{2}} + (v + v_{g}) \dfrac{dv}{dz} - \dfrac{q_{g}(h_{0} + v T_{0})^{2}}{(v + v_{g}) - q_{g}(h_{0} + v T_{0})} \left|\frac{dv}{dz} \right| \frac{dv}{dz} = \frac{1}{T} [v_{e}(v) - v]
\end{equation}
\\
\noindent
Making the analogy with the one-particle motion description, we can identify the friction coefficients and the pseudo-force that correspond to this model

\begin{equation*}
d_{1} = - \dfrac{q_{g}}{n}, \quad d_{2}= \biggl(\dfrac{q_{g}}{n(v + v_{g})}\biggr) \dfrac{q_{g}(h_{0} + v T_{0})^{2}}{(v + v_{g}) - q_{g}(h_{0} + v T_{0})}
\end{equation*}

\begin{equation*}
f = - \biggl(\dfrac{q_{g}}{n(v + v_{g})}\biggr)\frac{[v_{e}(v) - v]}{T} 
\end{equation*}
\\
\noindent
Where the dimensionless constants $(n, h_{0}, T_{0}, T)$, and the dimensionless fundamental diagram $v_{e}(v)$ are defined as follows

\begin{equation*}
n=\dfrac{\eta}{V_{max}}, \qquad h_{0}= \dfrac{H_{0}}{\rho_{max}}, \qquad T_{0} = (\rho_{max}V_{max})\mathcal{T},
\end{equation*}

\begin{equation*}
T = (\rho_{max}V_{max})\tau, \qquad v_{e}(v) = \dfrac{V_{e}(V)}{V_{max}}
\end{equation*}

An analytical analysis cannot be followed through due to the nonlinearity of Eq. (\ref{Eq8pollo}); as such, we proceed with a qualitative description of the model by means of a phase-space analysis. By introducing the next change of variable, the second order differential equation is transformed into a system of two first order differential equations

\begin{equation}\label{variablechangeB}
 \frac{dv}{dz} = c_{1}=y
\end{equation}

\begin{multline}\label{ecB}
\frac{dy}{dz} = c_{2} = \biggl( \dfrac{q_{g}}{n(v + v_{g})} \biggr) \biggl( (v + v_{g})y - \dfrac{q_{g}(h_{0} + v T_{0})^{2}}{(v + v_{g}) - q_{g}(h_{0} + v T_{0})} \left|y \right| y \\
- \dfrac{[v_{e}(v) - v]}{T} \biggr)
\end{multline}

With the aim to construct the phase space, the critical points need to be calculated, which are given by the condition $ \frac{dv}{dz}=0$. This is equivalent to computing

\begin{equation}\label{EPBKK}
[v_{e}(v_{c}) - v_{c}] = 0
\end{equation}
\\


\begin{table}[ht]
\begin{center}
\scalebox{0.80}{
\begin{tabular}{|cc||cc|}
\hline
$q_{g}$ $Q_{g}$ & $v_{g}$ $V_{g}$ & $v_{1}$  $V_{1}$ &  $v_{2}$  $V_{2}$ \\ \hline \hline
    
0.0952 & 0.11 & 0.0163 & 0.9063  \\ [0.2cm]
1600 veh/h & 12 $km/h$ & 0.63 $km/h$  & 108.63 $km/h$ \\[0.2cm] & & unstable spiral & saddle point \\[0.1cm] \hline
    
0.21 & 0.15 & 0.0776 & 0.7723  \\ [0.2cm]
3528 veh/h & 25.2 $km/h$ & 9.31 $km/h$ & 92.67 $km/h$ \\[0.2cm] & & unstable spiral & saddle point \\[0.1cm] \hline

0.0952 & -0.11 & 0.2343 & 0.8756 \\[0.2cm] 
1600 veh/h & -13.2 $km/h$ & 28.11 $km/h$ & 105.07 $km/h$ \\[0.2cm] & & unstable spiral & saddle point \\[0.1cm] \hline

0.0714 & -0.16 & 0.2559 & 0.9040 \\[0.2cm] 
1200 veh/h & -20 $km/h$ & 30.70 $km/h$ & 108.48 $km/h$ \\[0.2cm] & & unstable spiral & saddle point \\[0.1cm] \hline

0.0952 & 0.0 & 0.1065 & 0.8934  \\[0.2cm] 
1600 veh/h & 0 $km/h$ & 12.78 $km/h$ & 107.20 $km/h$ \\[0.2cm] & & unstable spiral & saddle point \\[0.1cm] \hline
\end{tabular}}
\label{tab:1}
\caption{Examples of critical values $v_c$ for the BKK model using the Greenshields diagram.}
\end{center}
\end{table}

\begin{table}[ht]
\begin{center} 
\scalebox{0.80}{
\begin{tabular}{ |cc||ccc| }
\hline
$q_{g}$ $Q_{g}$ & $v_{g}$ $V_{g}$ & $v_{1}$  $V_{1}$ & $v_{2}$  $V_{2}$ &  $v_{3}$  $V_{3}$ \\ \hline \hline
    
0.0952 & 0.11 & 0.1636 & 0.9337 & 3.15\e{-5} \\ [0.2cm]
1600 veh/h & 12 $km/h$ & 19.63 $km/h$ & 112 $km/h$ &  3.78\e{-3} $km/h$ \\[0.2cm] & & unstable spiral & saddle point & saddle point \\[0.1cm] \hline
    
0.15 & 0.21 & 0.2783 & 0.8624 & 0.0004  \\ [0.2cm]
2520 veh/h & 25.2 $km/h$ & 33.39 $km/h$ & 103.5 $km/h$ &  0.05 $km/h$ \\[0.2cm] & & unstable spiral & saddle point & saddle point \\[0.1cm] \hline

0.0952 & -0.11 & 0.4856 & 0.8953 &  \\[0.2cm] 
1600 veh/h & -13.2 $km/h$ & 58.27 $km/h$ & 107.4 $km/h$ & \\[0.2cm] & & unstable spiral & saddle point & \\[0.1cm] \hline

0.0714 & -0.16 & 0.4266 & 0.9325 &  \\[0.2cm] 
1200 veh/h & -20 $km/h$ & 52.24 $km/h$ & 112 $km/h$ & \\[0.2cm] & & unstable spiral & saddle point & \\[0.1cm] \hline

0.0952 & 0.0 & 0.3235 & 0.9199 &  \\[0.2cm] 
1600 veh/h & 0 $km/h$ & 8.16 $km/h$  & 113 $km/h$ & \\[0.2cm] & & unstable spiral & saddle point & \\[0.1cm] \hline
\end{tabular}}
\label{tab:2}
\caption{Examples of critical values $v_{c}$ for the BKK model using the Kerner-Konh\"{a}user diagram.}
\end{center}
\end{table}


When the Greenshields fundamental diagram, see Eq.(\ref{Greenshields}), is introduced in Eq.(\ref{EPBKK}), a concise expression is obtained for the critical points, which clearly depend on the values of the parameters $(q_{g}, v_{g})$

\begin{equation}\label{EPGreenshields1}
v_c = \dfrac{1}{2} \biggl( 1 - v_{g}\pm \sqrt{(1 - v_{g})^{2} - 4q_{g}} \biggr)
\end{equation}
\\

To determine the stability of the critical points, the eigenvalues of the corresponding Jacobian are calculated. Since the Jacobian elements depend on the parameters of the model and on the solutions $v_{c}$, it is expected to obtain different types of stability for different parameters. Clearly, an explicit calculation requires numerical values for all parameters, including the fundamental diagram. In this case, the following values are taken into account: $h_{0}=1$, $t_{0}=3.5$, $n=5$, and $T=140$.\\

Some examples of critical points $v_{c}$ for specific values of $q_g$ and $v_g$, and their associated stability are presented in Table 1 for the Greenshields diagram. On the other hand, when the Kerner-Konh\"{a}user fundamental diagram is introduced, see Eq.(\ref{KernerK}), several equilibrium points are produced. Likewise, some examples of critical points and their stability are given in Table 2.\\

Even though the numerical results cited in Table 1 and Table 2 provide insight into the behavior of the dynamical system, some simulations and qualitative analysis are presented in Sect. \ref{subsec:1} for specific trajectories in the phase space.


\section{The modified Helbing model}
\label{sec:4}

As in the case of the BKK model, the Helbing model proposes a continuity equation without sources for the density of vehicles $\rho(x,t)$, see Eq.(\ref{Eq4RMV}), a Navier-Stokes type of equation for the vehicles' flow velocity $V(x,t)$, and the velocity variance of vehicles $\Theta(x,t)$, see \cite{helbing95},

\begin{equation}\label{Eq2monis}
\frac{\partial (\rho V)}{\partial t} + \frac{\partial}{\partial x}(\rho V^{2} + \mathcal{P}) - \frac{\rho}{\tau} [V_{e}(V) - V] = 0
\end{equation}

\begin{equation}\label{Eq3monis}
\frac{\partial \Theta}{\partial t} +  V \frac{\partial \Theta}{\partial x} +
\frac{2 \mathcal{P}}{\rho} \frac{\partial V}{\partial x} + \frac{1}{\rho} \frac{\partial \mathcal{J}}{\partial x} - \frac{2}{\tau} \bigl[\Theta_{e} (V) - \Theta \bigr] = 0
\end{equation}
\\
\noindent
For this model, the term $\mathcal{P}$ resembles the pressure of traffic flow, wherein $\eta$ denotes a viscosity coefficient. The source term $S$ for this model is given by the fundamental diagram scaled by the flow density $\rho$ and a relaxation time $\tau$ that corresponds to the average acceleration time. 

\begin{equation*}\label{Eq4helbing}
\mathcal{P}(x,t) = \rho(x,t)\Theta(x,t) - \eta\dfrac{\partial V}{\partial x}
\end{equation*}

\begin{equation*}
S(x,t)= \frac{\rho}{\tau} [V_{e}(V) - V].
\end{equation*}

\noindent
Instead of considering the variance's dynamics as the Helbing model does, the equation system will be reduced by considering the following expression for the variance

\begin{equation}\label{theta}
\Theta = A(\rho)V^{2}
\end{equation}

\noindent where $A(\rho)$, which is known as the variance prefactor, is an empirical relation that relates the vehicle flow density with the variance's stationary state. Considering the new expression for the variance, the equations for the modified version of the Helbing model are as follows

\begin{equation}\label{Eq1Helbing2}
\frac{\partial \rho}{\partial t} + \frac{\partial (\rho V)}{\partial x} = 0
\end{equation}

\begin{equation}\label{Eq2Helbing2}
\frac{\partial V}{\partial t} +  V \frac{\partial V}{\partial x} + \frac{1}{\rho} \frac{\partial}{\partial x} \biggr( \rho A(\rho)V^{2} - \eta \dfrac{\partial V}{\partial x} \biggl) = \frac{1}{\tau} [V_{e}(V) - V].
\end{equation}
\\

As before, if we proceed with the application of the change of variable and the substitution of the dimensionless variables, the dimensionless equation for the velocity is obtained

\begin{equation}\label{1Helbing_simple}
- \biggl(\dfrac{n(v+v_{g})}{q_{g}}\biggr)\dfrac{d^{2}v}{dz^{2}} + \biggl((v+v_{g}) - \dfrac{A(\rho)v^{2}}{(v+v_{g})}\biggr)\dfrac{dv}{dz} + \dfrac{d}{dz} \bigr( A(\rho)v^{2} \bigl) = \dfrac{1}{T}[v_{e}(v)-v]
\end{equation}
\\
\noindent
Making the analogy with the one-particle motion description for this model, we can identify the friction coefficients and the pseudo-force term. It is important to notice that in this case, the nonlinear friction term $d_{2}$ is not present, because there are not any quadratic terms {\footnotesize $\biggl( \dfrac{dv}{dz}\biggr)^{2}$}. In relation to the linear friction coefficient $d_{1}$, the coefficient has a much more complicated structure, such that the explicit function in terms of $v$ and the parameters is not offered here

\begin{equation*}
d_{1} = d_{1}(v; v_{g}, q_{g}, A(\rho)), \quad d_{2}= 0
\end{equation*}

\begin{equation*}
f = - \biggl(\dfrac{q_{g}}{n(v + v_{g})}\biggr)\frac{[v_{e}(v) - v]}{T}. 
\end{equation*}
\\
\noindent
The nonlinearity of Eq.(\ref{1Helbing_simple}) prevents us from solving the system in an analytic way, therefore requiring we repeat the phase space analysis to propose a qualitative picture of the model's behavior. Through the next change of variable, a system of two first order equations is obtained

\begin{equation}\label{3-4Helbing_simple}
 \frac{dv}{dz} = c_{1} = y
\end{equation}

\begin{multline}\label{1.1Helbing_simple}
\dfrac{dy}{dz} = c_{2} = \biggl(\dfrac{q_{g}}{n} - \dfrac{A(\rho)v^{2} q_{g}}{n(v+v_{g})^{2}}\biggr)y + \biggl( \dfrac{q_{g}}{n(v + v_{g})} \biggr) \biggl( 2A(\rho)vy + v^{2}\dfrac{dA(\rho)}{dz} \biggr)\\
 - \dfrac{q_{g}[v_{e}(v) - v]}{nT(v + v_{g})}
\end{multline}

This autonomous dynamical system, described by the variables $(v, y)$ and the parameters $(v_{g}, q_{g}, n, T)$, is now comparable to the system obtained for the modified BKK model, so a comparison in the phase space is now possible.\\

As for the prefactor $A(\rho)$, the following expression is applied in the calculation of the critical points

\begin{equation}
A(\rho) = \left(A_{0} + \Delta A \left[\tanh \left[\frac{\frac{q_{g}}{v + v_{g}}-\rho_{c}}{\Delta \rho}\right] + 1\right]\right),
\end{equation}
\\
\noindent
where the values considered are: $A_{0} = 0.008$, $\Delta A= 0.015$, $\rho_{c}=0.28$, $\Delta \rho=0.1$, $n=5$ and $T=140$. Therefore, the critical points $(v_c)$ corresponding to Eqs. (\ref{3-4Helbing_simple}, \ref{1.1Helbing_simple}) are obtained from the condition $y=0$, equivalent to solving

\begin{equation}\label{EP1.2Helbing}
[v_{e}(v_{c})- v_{c}] = 0.
\end{equation}

\noindent
It is important to mention that the critical points $v_c$ corresponding to this modified version of the Helbing model are equal to the values that are obtained for the full Helbing model; i.e they are independent of the variance's dynamic because the same condition $[v_{e}(v_{c})- v_{c}] = 0$ applies to both cases. The same happens with the stability of the critical points. Although the eigenvalues of the Jacobian corresponding to the linearized system written in Eqs.(\ref{3-4Helbing_simple}, \ref{1.1Helbing_simple}) are different from the eigenvalues associated to the original Helbing model, they preserve the same type of stability.\\

It is clear that the coordinates of the critical points depend on the specific expressions chosen for the fundamental diagram $v_{e}$. As before, if the Greenshields fundamental diagram, see Eq.(\ref{Greenshields}), is employed in Eq.(\ref{EP1.2Helbing}), the critical points $v_{c}$ are determined again by

\begin{equation}\label{EPGreenshields}
v_c = \dfrac{1}{2} \biggl( 1 - v_{g}\pm \sqrt{(1 - v_{g})^{2} - 4q_{g}} \biggr).
\end{equation} 


\begin{table}[ht]
\begin{center}
\scalebox{0.75}{
\begin{tabular}{|cc||cc|}
\hline
$q_{g}$ $Q_{g}$ & $v_{g}$ $V_{g}$ & $v_{1}$  $V_{1}$ &  $v_{2}$  $V_{2}$ \\ \hline \hline
    
0.0952 & 0.11 & 0.0163 & 0.9063  \\ [0.2cm]
1600 veh/h & 12 $km/h$ & 0.63 $km/h$  & 108.63 $km/h$ \\[0.2cm] & & unstable spiral & saddle point \\[0.1cm] \hline
    
0.21 & 0.15 & 0.0776 & 0.7723  \\ [0.2cm]
3528 veh/h & 25.2 $km/h$ & 9.31 $km/h$ & 92.67 $km/h$ \\[0.2cm] & & unstable spiral & saddle point \\[0.1cm] \hline

0.0952 & -0.11 & 0.2343 & 0.8756 \\[0.2cm] 
1600 veh/h & -13.2 $km/h$ & 28.11 $km/h$ & 105.07 $km/h$ \\[0.2cm] & & unstable spiral & saddle point \\[0.1cm] \hline

0.0714 & -0.16 & 0.2559 & 0.9040 \\[0.2cm] 
1200 veh/h & -20 $km/h$ & 30.70 $km/h$ & 108.48 $km/h$ \\[0.2cm] & & unstable spiral & saddle point \\[0.1cm] \hline

0.0952 & 0.0 & 0.1065 & 0.8934  \\[0.2cm] 
1600 veh/h & 0 $km/h$ & 12.78 $km/h$ & 107.20 $km/h$ \\[0.2cm] & & unstable spiral & saddle point \\[0.1cm] \hline
\end{tabular}}
\label{table3}
\caption{Examples of critical values $v_c$ for the Helbing model using the Greenshields diagram.}
\end{center}
\end{table}

\begin{table}[ht]
\begin{center} 
\scalebox{0.75}{
\begin{tabular}{|cc||ccc|}
\hline
$q_{g}$ $Q_{g}$ & $v_{g}$ $V_{g}$ & $v_{1}$  $V_{1}$ & $v_{2}$  $V_{2}$ &  $v_{3}$  $V_{3}$ \\ \hline \hline
    
0.0952 & 0.11 & 0.1636 & 0.9337 & 3.15\e{-5} \\ [0.2cm]
1600 veh/h & 12 $km/h$ & 19.63 $km/h$ & 112 $km/h$ &  3.78\e{-3} $km/h$ \\[0.2cm] & & unstable spiral & saddle point & saddle point \\[0.1cm] \hline
    
0.15 & 0.21 & 0.2783 & 0.8624 & 0.0004  \\ [0.2cm]
2520 veh/h & 25.2 $km/h$ & 33.39 $km/h$ & 103.5 $km/h$ &  0.05 $km/h$ \\[0.2cm] & & unstable spiral & saddle point & saddle point \\[0.1cm] \hline

0.0952 & -0.11 & 0.4856 & 0.8953 &  \\[0.2cm] 
1600 veh/h & -13.2 $km/h$ & 58.27 $km/h$ & 107.4 $km/h$ & \\[0.2cm] & & unstable spiral & saddle point & \\[0.1cm] \hline

0.0714 & -0.16 & 0.4266 & 0.9325 &  \\[0.2cm] 
1200 veh/h & -20 $km/h$ & 52.24 $km/h$ & 112 $km/h$ & \\[0.2cm] & & unstable spiral & saddle point & \\[0.1cm] \hline

0.0952 & 0.0 & 0.3235 & 0.9199 &  \\[0.2cm] 
1600 veh/h & 0 $km/h$ & 8.16 $km/h$  & 113 $km/h$ & \\[0.2cm] & & unstable spiral & saddle point & \\[0.1cm] \hline
\end{tabular}}
\label{table4}
\caption{Examples of critical values $v_{c}$ for the Helbing model using the Kerner-Konh\"{a}user diagram.}
\end{center}
\end{table}


\noindent
On the other hand, when the Kerner-Konh\"{a}user fundamental diagram, Eq.(\ref{KernerK}), is introduced, two equilibrium points are produced plus one extra critical point in the vicinity of zero for some values of $q_g$ and $v_g$.\\

To study the stability of the critical points, the eigenvalues of the Jacobian corresponding to the linearized system written in Eqs.(\ref{3-4Helbing_simple}, \ref{1.1Helbing_simple}) are calculated. Through an immediate calculation, it is shown that the elements of the Jacobian matrix, see Eq.(\ref{jacobian}), depend on the parameters $q_g$ and $v_g$, and on the solutions for the critical points $v_{c}$. Therefore, for different values it is expected to find different kinds of stability. Some examples of critical points and their stability are given in Table 3 for the Greenshields diagram, and in Table 4 for the Kerner-Konh\"{a}user diagram.


\section{Trajectories in the phase space}
\label{sec:5}

A qualitative analysis alone will be considered to study the behavior of both models. For each one, only three of the five cases presented in the tables of the previous sections will be considered for different $v_{g}$ and $q_{g}$ values, for which the stability of the critical points will be analyzed along with their respective phase diagrams.

\subsection{Phase planes of the modified BKK model}
\label{subsec:1}

In Table 1, the specific values of the equilibrium points $v_{c}$ are shown by evaluating Eq. (\ref{variablechangeB}) and Eq. (\ref{ecB}) with the chosen parameters. It is important to notice here how the $v_{c}$ values are true for the Greenshields diagram, while the values found in Table 2 are pertinent to the Kerner-Konh\"{a}user diagram.\\

\begin{figure}[H]
	\includegraphics[width=\textwidth]{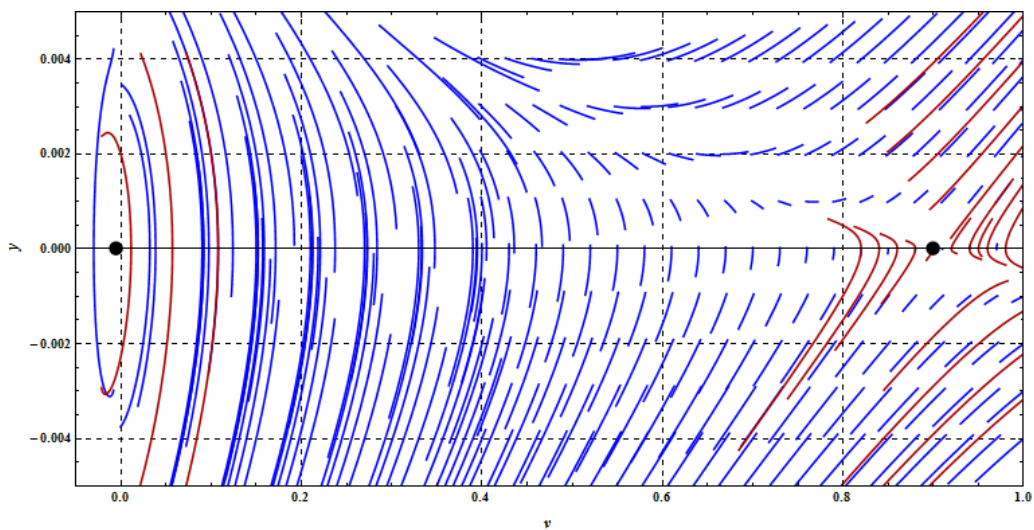}
	\centering
	\caption{Phase plane of the BKK model using the Greenshields diagram, with $q_{g}=0.0952$ and $v_{g}=0.11$.}
	\label{fig1}
\end{figure}

The procedure to determine the stability of the critical points $v_{c}$ by finding the eigenvalues of the Jacobian matrix is then applied. Going back to Table 1, for the first pair of values $v_{g}$ and $q_{g}$, the two eigenvalues for $v_{1}$ are complex, with both real parts positive. Thus, an unstable spiral is obtained for the first equilibrium point. For $v_{2}$, one eigenvalue is positive while the other is negative, yielding a saddle point. The phase plane is then determined by equations Eq.(\ref{variablechangeB}) and Eq.(\ref{ecB}), from which the trajectories are drawn, see Fig. \ref{fig1}.\\

\begin{figure}[H]
	\includegraphics[width=\textwidth]{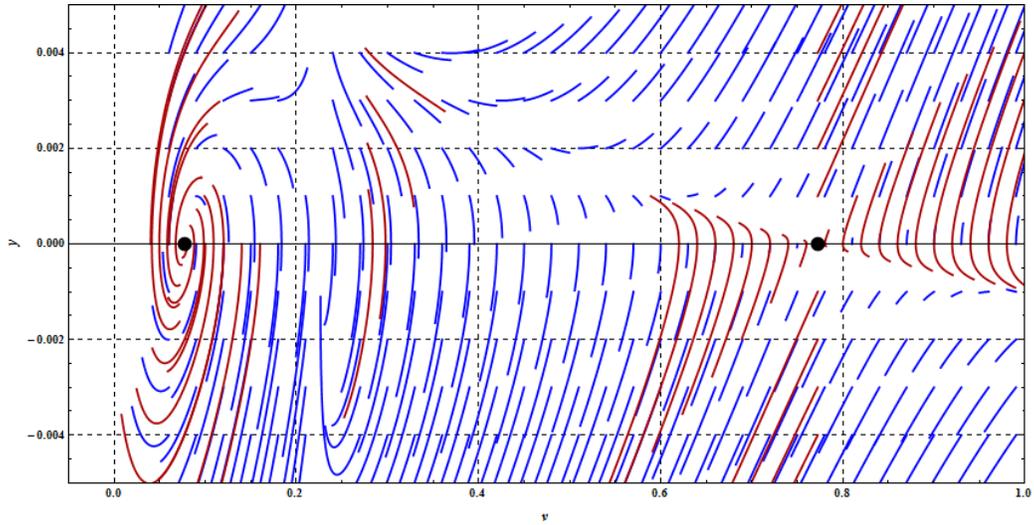}
	\centering
	\caption{Phase plane of the BKK model using the Greenshields diagram, with $q_{g}=0.21$ and $v_{g}=0.15$.}
	\label{fig2}
\end{figure}

For the second pair of values $v_{g}$ and $q_{g}$, the eigenvalues offer the same stability as the previous ones, for which an unstable spiral and a saddle point are obtained. Again, drawing the phase plane from the system of equations Eq.(\ref{variablechangeB}) and Eq.(\ref{ecB}), the phase plane is obtained, see Fig. \ref{fig2}\\
 
\begin{figure}[H]
	\includegraphics[width=\textwidth]{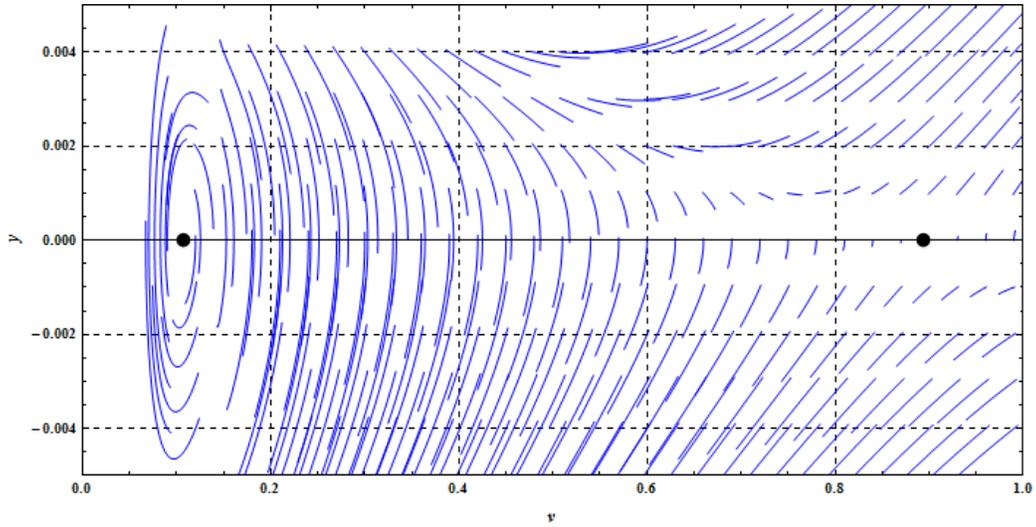}
	\centering
	\caption{Phase plane of the BKK model using the Greenshields diagram, with $q_{g}=0.0952$ and $v_{g}=0$.}
	\label{fig3}
\end{figure}

For the fifth pair of $v_{g}$ and $q_{g}$ values, the equilibrium points turn out to be unstable; an unstable spiral and a saddle point are once again the result, see Fig. \ref{fig3}\\

\begin{figure}[H]
	\includegraphics[width=\textwidth]{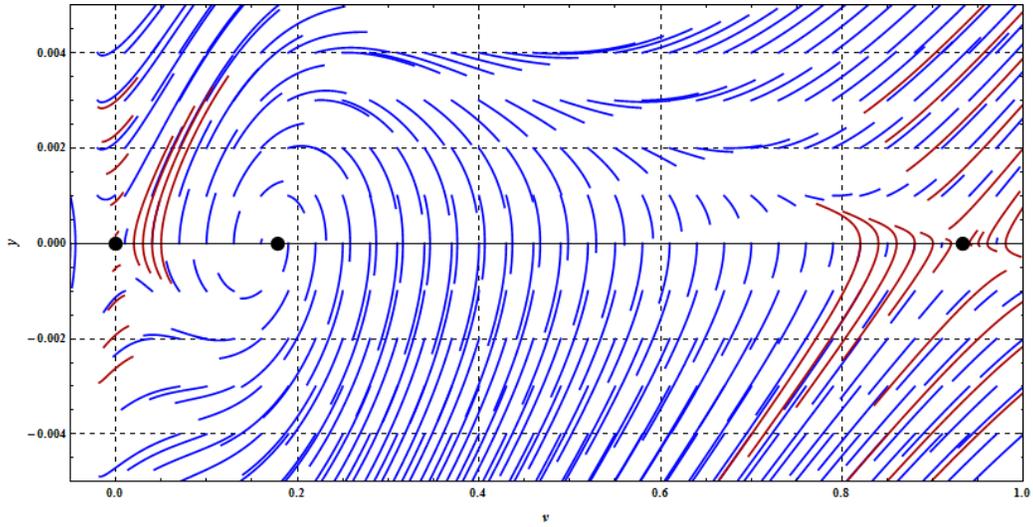}
	\centering
	\caption{Phase plane of the BKK model using the Kerner-Konh\"{a}user diagram, with $q_{g}=0.0952$ and $v_{g}=0.11$.}
	\label{fig4}
\end{figure}

As seen in Table 2, now there are three equilibrium points for the first and second pair of values $v_{g}$ and $q_{g}$ when the Kerner-Konh\"{a}user diagram is introduced in Eq.(\ref{EPBKK}). The analysis of the eigenvalues is analogous to the previous one using the Greenshields diagram. For the first pair of values, the equlibrium points consist of an unstable spiral and two saddle points, see Fig. \ref{fig4}.\\

\begin{figure}[H]
	\includegraphics[width=\textwidth]{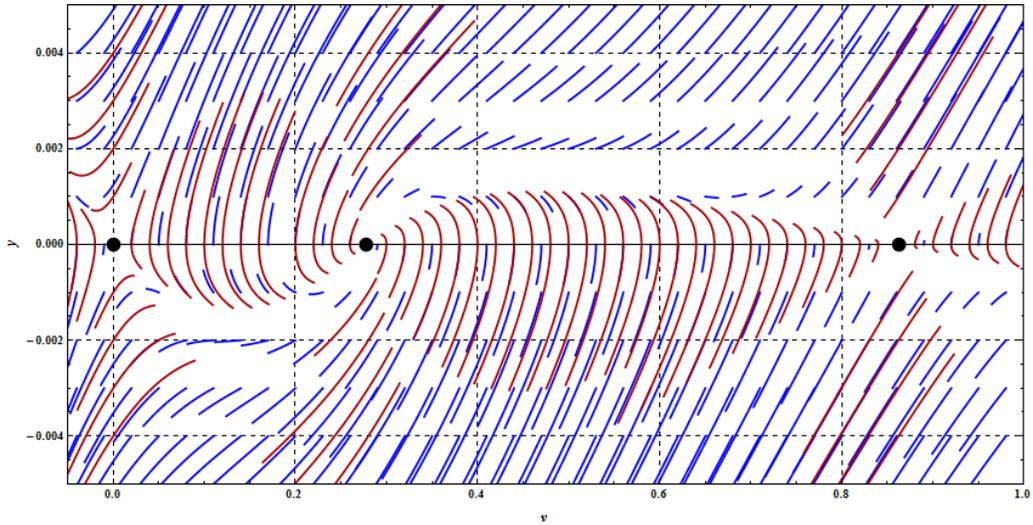}
	\centering
	\caption{Phase plane of the BKK model using the Kerner-Konh\"{a}user diagram, with $q_{g}=0.15$ and $v_{g}=0.21$.}
	\label{fig5}
\end{figure}

Similarly, for the second pair of values the equilibrium points consist of two saddle points and one unstable spiral, see Fig. \ref{fig5}.\\

\begin{figure}[H]
	\includegraphics[width=\textwidth]{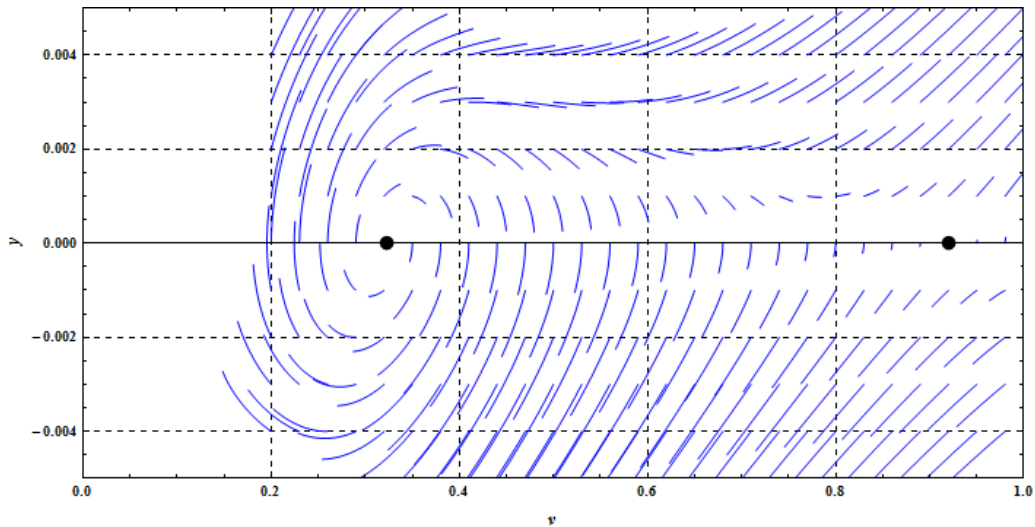}
	\centering
	\caption{Phase plane of the BKK model using the Kerner-Konh\"{a}user diagram, with $q_{g}=0.0952$ and $v_{g}=0$.}
	\label{fig6}
\end{figure}

And for the fifth pair of values, there are only two equilibrium points, an unstable spiral and a saddle point, see Fig. \ref{fig6}\\

\subsection{Phase planes of the modified Helbing model}
\label{subsec:2}

The critical points of the modified Helbing model are given by Eq.(\ref{EP1.2Helbing}). In a similar fashion, the Jacobian matrix is generated by using the system of equations composed by Eq.(\ref{3-4Helbing_simple}) and Eq.(\ref{1.1Helbing_simple}), from which the trajectories of the phase plane are drawn.\\ 

In Table 3, the equilibrium points for five instances of values $v_{g}$ and $q_{g}$ are shown using the Greenshields diagram, and in Table 4 the equilibrium points are shown using the Kerner-Konh\"{a}user diagram.\\

\begin{figure}[H]
	\includegraphics[width=\textwidth]{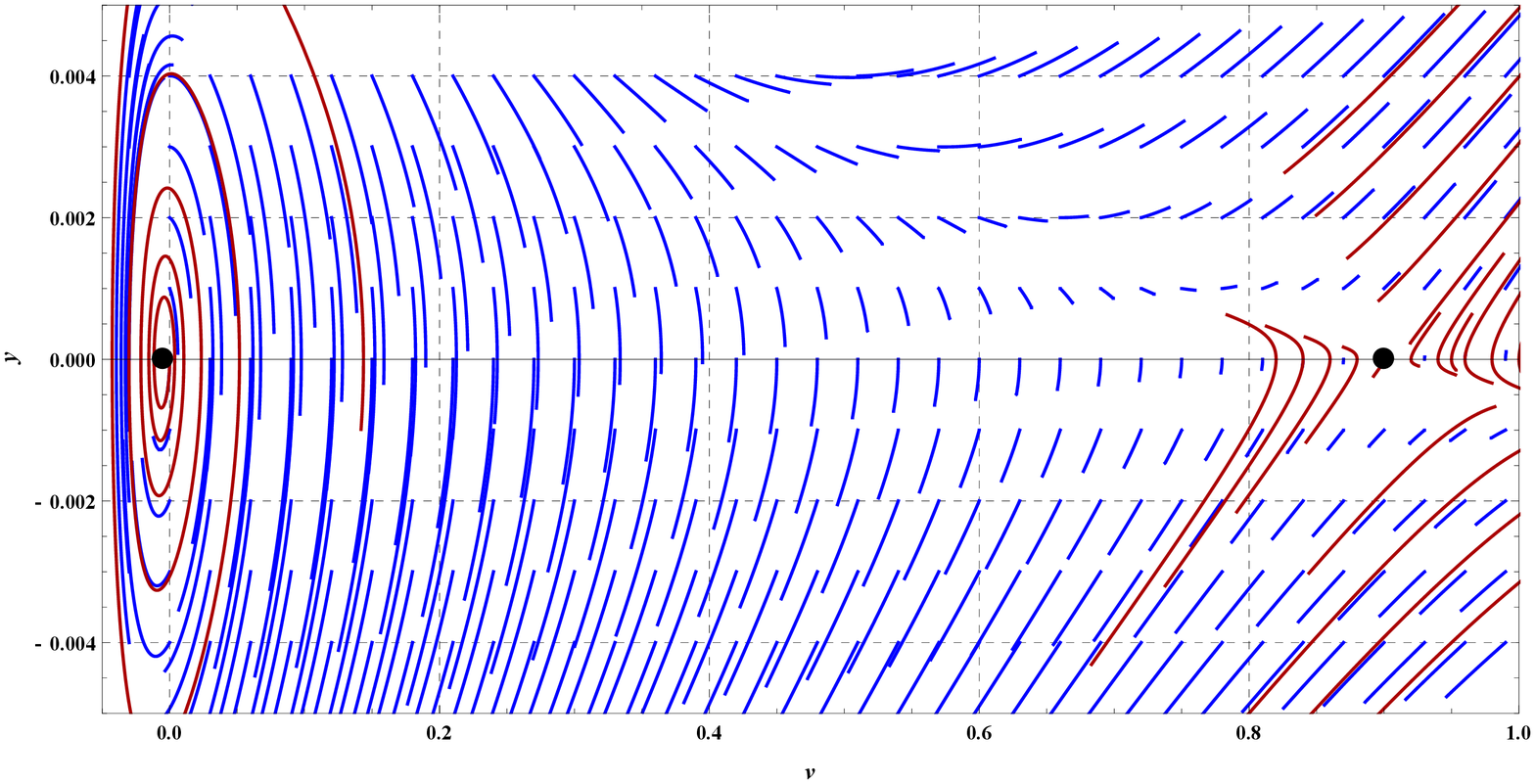}
	\centering
	\caption{Phase plane of the Helbing model using the Greenshields diagram, with $q_{g}=0.0952$ and $v_{g}=0.11$.}
	\label{fig7}
\end{figure}

From Table 3, for the first pair of values $v_{g}$ and $q_{g}$, an unstable spiral and a saddle point are the resulting equilibrium points, see Fig. \ref{fig7}.\\

\begin{figure}[H]
	\includegraphics[width=\textwidth]{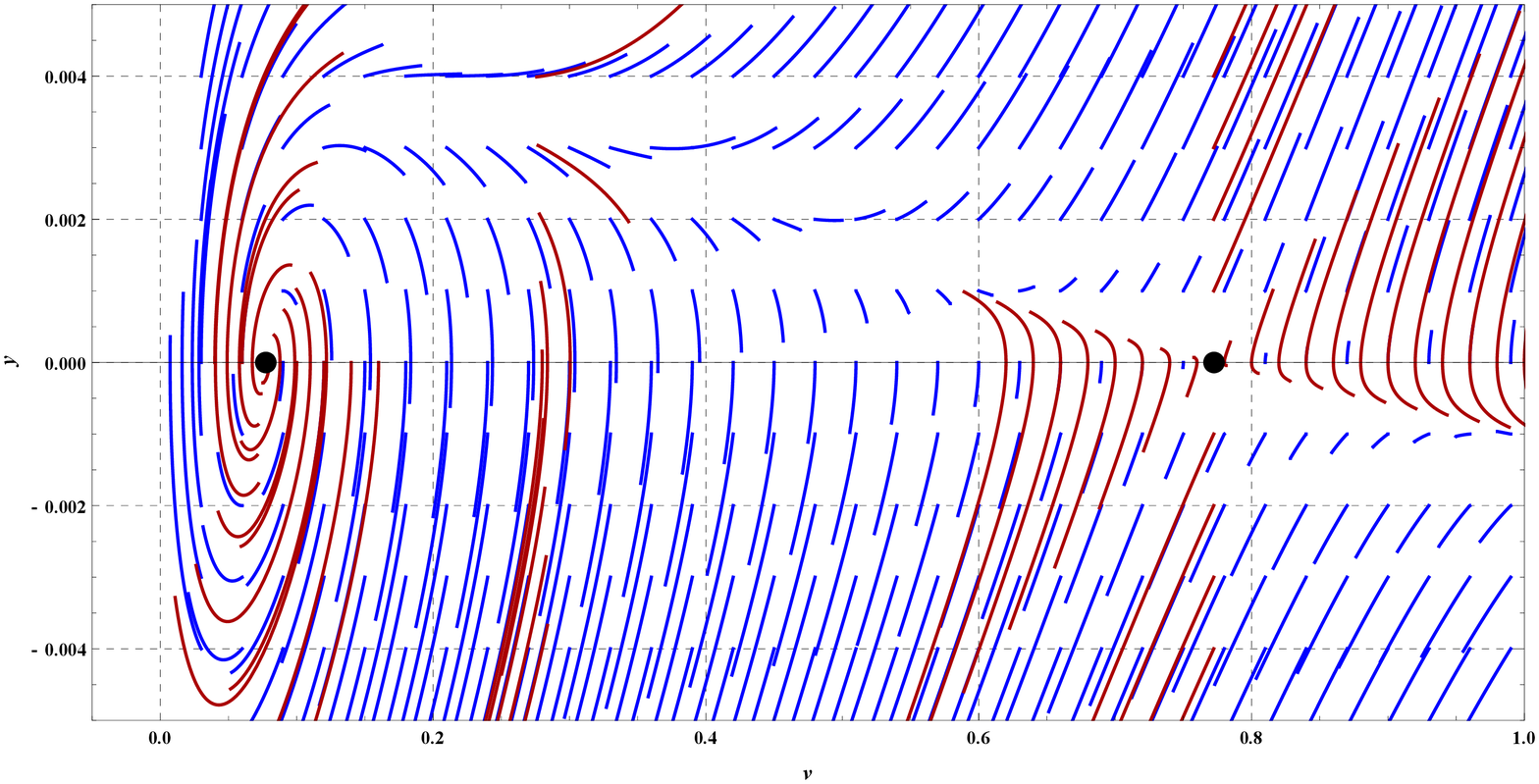}
	\centering
	\caption{Phase plane of the Helbing model using the Greenshields diagram, with $q_{g}=0.21$ and $v_{g}=0.15$.}
	\label{fig8}
\end{figure}

For the second pair of values, its corresponding eigenvalues are calculated, from which the equilibrium points are an unstable spiral and a saddle point, see Fig. \ref{fig8}.\\

\begin{figure}[H]
	\includegraphics[width=\textwidth]{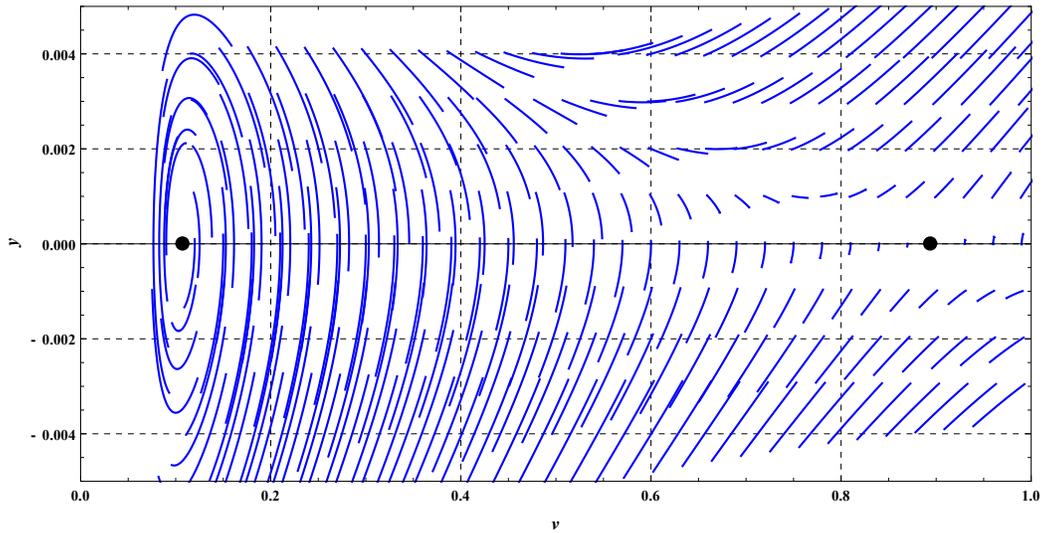}
	\centering
	\caption{Phase plane of the Helbing model using the Greenshields diagram, with $q_{g}=0.0952$ and $v_{g}=0$.}
	\label{fig9}
\end{figure}

The fifth pair of values equally yield two equilibrium points, thereby producing an unstable spiral and a saddle point, see Fig. \ref{fig9}.\\

\begin{figure}[H]
	\includegraphics[width=\textwidth]{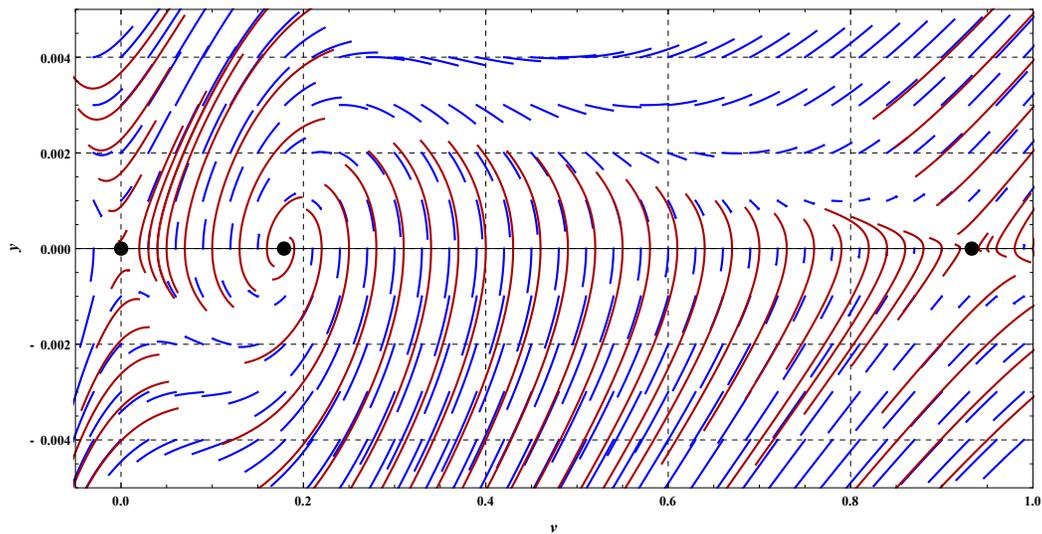}
	\centering
	\caption{Phase plane of the Helbing model using the Kerner-Konh\"{a}user diagram, with $q_{g}=0.0952$ and $v_{g}=0.11$.}
	\label{fig10}
\end{figure}

Following from the Kerner-Konh\"{a}auser examples in the modified BKK model, it is noted from Table 4 that there are three equilibrium points for the first and second pair of values $v_{g}$ and $q_{g}$, when introducing the diagram in equation Eq.(\ref{EP1.2Helbing}). Therefore, for the first pair of values there are three corresponding equilibrium points, an unstable spiral and two saddle points, see Fig. \ref{fig10}.\\

\begin{figure}[H]
	\includegraphics[width=\textwidth]{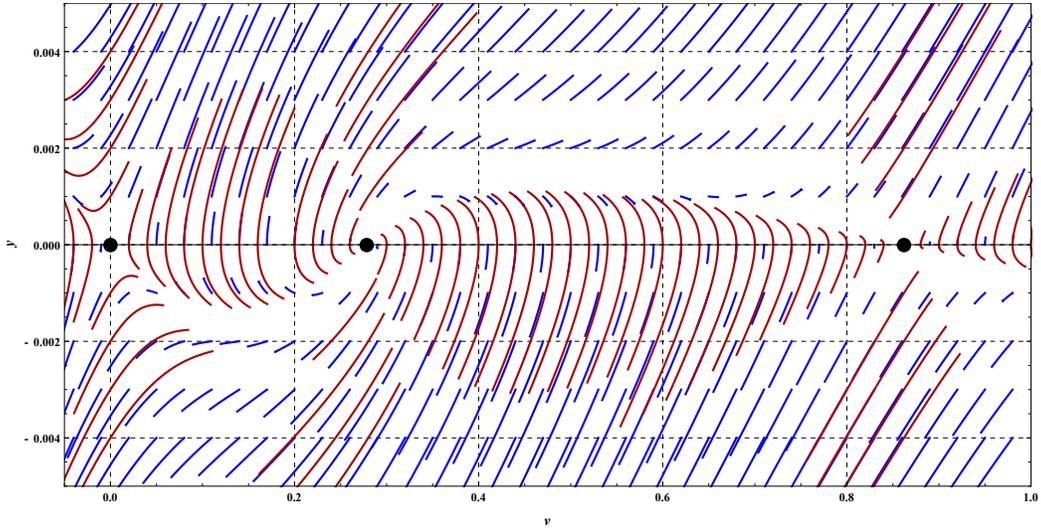}
	\centering
	\caption{Phase plane of the Helbing model using the Kerner-Konh\"{a}user diagram, with $q_{g}=0.15$ and $v_{g}=0.21$.}
	\label{fig11}
\end{figure}

Equally, from the second pair of values, three equilibrium points are obtained; two saddle points and an unstable spiral. See Fig. \ref{fig11}.\\

\begin{figure}[H]
	\includegraphics[width=\textwidth]{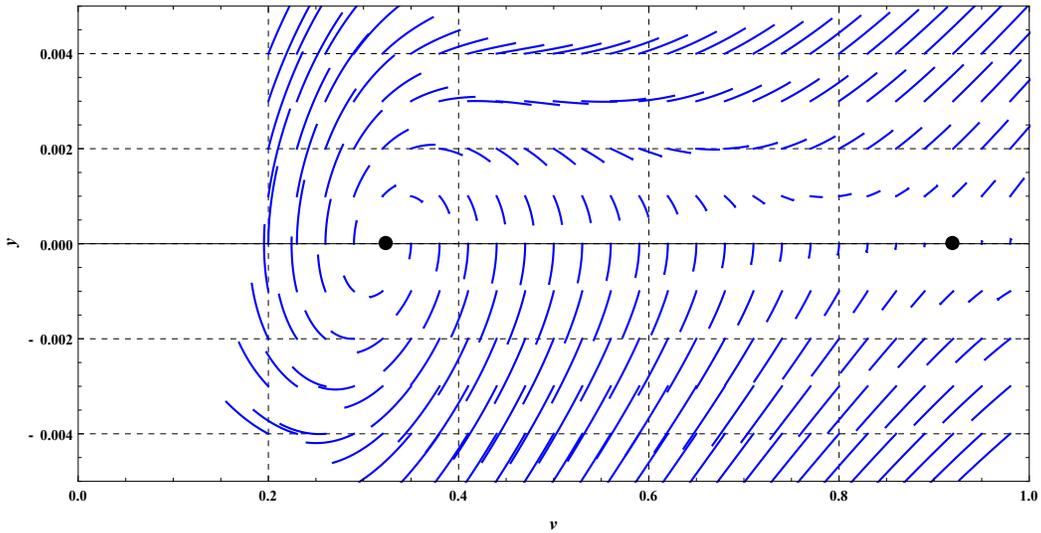}
	\centering
	\caption{Phase plane of the Helbing model using the Kerner-Konh\"{a}user diagram, with $q_{g}=0.0952$ and $v_{g}=0$.}
	\label{fig12}
\end{figure}

For the fifth pair of values, there are only two equilibrium points; an unstable spiral and a saddle point. See Fig. \ref{fig12}.\\

\section{Conclusions}
\label{sec:6}

\begin{itemize}
	\item 
	The critical points found via the Jacobian matrix are equal in both models for the same parameters $q_{g}$ and $v_{g}$. This is due to the fact that the evaluation of the critical points depends only on the fundamental diagram and is independent of other dynamic terms. Local stability is the same as a consequence, thus obtaining only two kinds of critical points: saddle points and unstable spirals.\\
\end{itemize}

\begin{itemize}
	\item 
	Examination of the phase diagrams supports the local stability analysis, since the trajectories for different parameters show an equal behavior around the same critical points. It can be noted then, that the structure of each model's equations has no consequences in terms of the local behavior around equilibrium points; different parameters like the relaxation time in the Helbing model and the braking times in the BKK model do not significantly affect in this respect. Also, the analysis performed in the phase space indicates that no heteroclinic trajectories can be found in any of the cases presented.\\
\end{itemize}

\begin{itemize}
	\item 
	Models with a similar structure, following a Navier-Stokes type of equation to approach the structure of the Kerner-Konh\"{a}user model with the velocity as sole dynamic variable, will tend to share the same behavior around equilibrium points considering that ultimately, they will only depend on the fundamental diagram regardless of the other factors taken into account, as studied with these two models.\\
\end{itemize}


\section*{References}





\end{document}